\documentclass[12pt]{article}

\newcommand{\sumjk}{\sum_{\scriptstyle j,k \atop \scriptstyle i\ne j\ne k\ne
i}}
\newcommand{\sumij}{\sum_{\scriptstyle i,j \atop \scriptstyle i\ne j}}
\newcommand{\ad}{a^{\vphantom{\dagger}}}
\newcommand{\ac}{a^{\dagger}}
\def\case#1#2{{\textstyle{#1\over #2}}}

\begin{document}

\baselineskip=22pt plus 1pt minus 1pt
%
%
\begin{center}

{\bf EXCHANGE OPERATORS AND EXTENDED HEISENBERG ALGEBRA FOR THE
THREE-BODY CALOGERO-MARCHIORO-WOLFES PROBLEM}\\[3cm]

C. QUESNE \footnote{Directeur de recherches FNRS}$^,$\footnote{E-mail:
cquesne@ulb.ac.be}\\

{\em Physique Nucl\'eaire Th\'eorique et Physique Math\'ematique,
Universit\'e Libre de Bruxelles, Campus de la Plaine CP229, Boulevard du
Triomphe, B-1050 Brussels, Belgium}\\[5cm]

\end{center}

\begin{abstract}

\noindent
The exchange operator formalism previously introduced for the Calogero problem
is extended to the three-body Calogero-Marchioro-Wolfes one. In the absence of
oscillator potential, the Hamiltonian of the latter is interpreted as a free
particle
Hamiltonian, expressed in terms of generalized momenta. In the presence of
oscillator potential, it is regarded as a free modified boson Hamiltonian. The
modified boson operators are shown to belong to a $D_6$-extended Heisenberg
algebra. A proof of complete integrability is also provided.

\end{abstract}

\clearpage
%
%
\section{Introduction}
A long time ago, Calogero~\cite{1} solved the Schr\H odinger equation for three
particles in one dimension, interacting pairwise via harmonic and inverse
square
potentials. Later, Wolfes~\cite{2} extended Calogero's method to the case where
there is an additional three-body potential of a special form. The same problem
in
the absence of harmonic potential was also studied by Calogero and
Marchioro~\cite{3}. Other exactly solvable many-body problems were then
introduced, analyzed from the viewpoint of classical or quantum integrability,
and
shown to be related to root systems of Lie algebras~\cite{4}.\par
%
%
A breakthrough in the study of integrable models occurred some three years ago
when Brink {\em et al}~\cite{5} and Polychronakos~\cite{6} independently
introduced
an exchange operator formalism, leading to covariant derivatives, known in the
mathematical literature as Dunkl operators~\cite{7}, and to an $S_N$-extended
Heisenberg algebra~\cite{8,9}. In terms of the latter, the $N$-body
quantum-mechanical Calogero model can indeed be interpreted as a model of free
modified oscillators. Such an approach emphasizes the relations between the
Calogero problem and fractional statistics~\cite{10} and is connected with the
spin generalization of the former (see e.g. Ref.~\cite{11} and references
quoted
therein).\par
%
%
Recently, there has been a renewed interest in the three-body
Calogero-Marchioro-Wolfes (CMW) problem~\cite{2,3} and some other related
three-particle problems including a three-body potential. Khare and
Bhaduri~\cite{12} indeed showed that they can be solved by using supersymmetric
quantum-mechanical techniques. Their approach is based upon Calogero's original
method~\cite{1}, wherein after eliminating the centre-of-mass motion, the
three-body problem is mapped to that of a particle on a plane and the
corresponding
Schr\H odinger equation is separated into a radial and an angular
equations.\par
%
%
The purpose of the present letter is to show that the exchange operator
formalism
of Brink {\em et al}~\cite{5} and Polychronakos~\cite{6} can be extended to the
CMW
problem, thereby leading to a further generalization of the Heisenberg algebra.
Our
starting point will be another supersymmetric approach to the three-body
problem,
wherein the latter is mapped to that of a particle in three-dimensional space
and
use is made of the Andrianov {\em et al} generalization of supersymmetric
quantum
mechanics for multidimensional Hamiltonians~\cite{13}.\par
%
%
\section{Supersymmetric Approach to the CMW Problem}
The three-particle Hamiltonian of the CMW problem is given by~\cite{2,3}
\begin{equation}
  H = \sum_{i=1}^3 \left(-\partial_i^2 + \omega^2 x_i^2\right)
       + g \sum_{\scriptstyle i,j=1 \atop \scriptstyle i\ne j}^3
\frac{1}{(x_i-x_j)^2}
       + 3f  \sum_{\scriptstyle i,j,k=1 \atop \scriptstyle i\ne j\ne k\ne i}^3
         \frac{1}{(x_i+x_j-2x_k)^2}, \label{eq:2.1}
\end{equation}
where $x_i$, $i=1$, 2,~3, denote the particle coordinates, $\partial_i \equiv
\partial/\partial x_i$, and the inequalities $g > -1/4$, and $f > -1/4$ are
assumed
to be valid to prevent collapse. Let $x_{ij} \equiv x_i - x_j$, $i\ne j$, and
$y_{ij}
\equiv x_i + x_j - 2x_k$, $i\ne j\ne k\ne i$, where in the latter we suppressed
index~$k$ as it is entirely determined by $i$ and~$j$.\par
%
%
Since for singular potentials crossing is not allowed, in the case of
distinguishable
particles the wave functions in different sectors of configuration space are
disconnected. We shall therefore restrict the particle coordinates to the
ranges
$x_1 > x_2 > x_3$ if $g\ne 0$, $f = 0$, $x_1 > x_3$, $x_2 > x_3$, $|x_1-x_2| <
x_1-x_3$, $|x_1-x_2| < x_2-x_3$ if $g = 0$, $f \ne 0$, and $x_1 > x_2 > x_3$,
$x_1-x_2 < x_2-x_3$ if $g\ne 0$, $f\ne 0$. In the case of indistinguishable
particles, there is an additional symmetry requirement, which we shall not
review
here as it was discussed in detail in Refs.~\cite{2} and~\cite{3}.\par
%
%
For distinguishable particles, the unnormalized ground-state wave function of
Hamiltonian~(\ref{eq:2.1}), corresponding to the eigenvalue
\begin{eqnarray}
  E_0
      & = & 3\omega (2\kappa + 1) \qquad \mbox{if }g\ne 0, f = 0, \nonumber \\
      & = & 3\omega (2\lambda + 1) \qquad \mbox{if }g = 0, f\ne 0, \nonumber \\
      & = & 3\omega (2\kappa + 2\lambda + 1) \qquad \mbox{if }g\ne 0, f\ne 0,
  \label{eq:2.1a}
\end{eqnarray}
is
\begin{eqnarray}
  \psi_0(\mathbf{x})
      & = & \exp \left(- \case{1}{2} \omega \sum_i x_i^2\right) |x_{12} x_{23}
            x_{31}|^{\kappa} \qquad \mbox{if }g\ne 0, f = 0, \nonumber \\
      & = & \exp \left(- \case{1}{2} \omega \sum_i x_i^2\right) |y_{12} y_{23}
            y_{31}|^{\lambda} \qquad \mbox{if }g = 0, f\ne 0, \nonumber \\
      & = & \exp \left(- \case{1}{2} \omega \sum_i x_i^2\right) |x_{12} x_{23}
            x_{31}|^{\kappa}  |y_{12} y_{23} y_{31}|^{\lambda} \qquad \mbox{if
}g\ne 0,
            f\ne 0,
  \label{eq:2.2}
\end{eqnarray}
where $\kappa \equiv \frac{1}{2} (1 + \sqrt{1 + 4g})$, $\lambda \equiv
\frac{1}{2}
(1 + \sqrt{1 + 4f})$ (implying $g = \kappa (\kappa - 1)$, $f = \lambda (\lambda
-
1)$). In terms of the function $\chi(\mathbf{x}) = - \ln \psi_0(\mathbf{x})$,
one
can construct six differential operators $Q_i^{\pm} = \mp \partial_i +
\partial_i
\chi$, $i=1$, 2,~3, whose explicit form is given by
\begin{eqnarray}
  Q_i^{\pm}
     & = & \mp \partial_i + \omega x_i - \kappa \sum_{j\ne i} \frac{1}{x_{ij}}
           \qquad \mbox{if }g\ne 0, f = 0, \nonumber \\
     & = &\mp \partial_i + \omega x_i - \lambda \left(\sum_{j\ne i}
\frac{1}{y_{ij}}
           - \sumjk \frac{1}{y_{jk}}\right) \qquad \mbox{if }g = 0, f\ne 0,
\nonumber
           \\
     & = & \mp \partial_i + \omega x_i - \kappa \sum_{j\ne i} \frac{1}{x_{ij}}
-
           \lambda \left(\sum_{j\ne i} \frac{1}{y_{ij}} - \sumjk
\frac{1}{y_{jk}}\right)
            \qquad \mbox{if }g\ne 0, f\ne 0.
  \label{eq:2.3}
\end{eqnarray}
\par
%
%
It can be easily shown~\cite{13} that $H - E_0$ can be regarded as the
$H^{(0)}$
component of a supersymmetric Hamiltonian
\begin{equation}
  \hat H = \left(
      \begin{array}{cccc}
           H^{(0)} & 0         & 0         & 0        \\
           0         & H^{(1)} & 0         & 0        \\
           0         & 0         & H^{(2)} & 0        \\
           0         & 0         & 0         & H^{(3)}
      \end{array}
   \right) \label{eq:2.4}
\end{equation}
with supercharge operators
\begin{equation}
  \hat Q^+ = \left(
      \begin{array}{cccc}
           0             & 0             & 0             & 0        \\
           Q^+_{0,1} & 0             & 0             & 0        \\
           0             & Q^+_{1,2} & 0             & 0        \\
           0             & 0             & Q^+_{2,3} & 0
      \end{array}
  \right), \qquad
  \hat Q^- = \left(\hat Q^+\right)^{\dagger} = \left(
      \begin{array}{cccc}
           0 & Q^-_{1,0} & 0             & 0                    \\
           0 & 0             & Q^-_{2,1} & 0                    \\
           0 & 0             & 0             & Q^-_{3,2}        \\
           0 & 0             & 0             & 0
      \end{array}
   \right),     \label{eq:2.5}
\end{equation}
i.e., $\hat H$, $\hat Q^+$, $\hat Q^-$ generate the supersymmetric
quantum-mechanical algebra sqm(2),
\begin{equation}
  \left\{\hat Q^+, \hat Q^-\right\} = \hat H, \qquad \left\{\hat Q^+, \hat
Q^+\right\}
  = \left\{\hat Q^-, \hat Q^-\right\} = 0, \qquad \left[\hat Q^+, \hat H\right]
=
  \left[\hat Q^-, \hat H\right] = 0.    \label{eq:2.6}
\end{equation}
In (\ref{eq:2.5}), $Q^+_{0,1}$, $Q^+_{1,2}$, $Q^+_{2,3}$ (resp.~$Q^-_{1,0}$,
$Q^-_{2,1}$, $Q^-_{3,2}$) denote $3\times1$, $3\times3$, $1\times3$
(resp.~$1\times3$, $3\times3$, $3\times1$) matrices, whose elements are
$Q^-_i$, $P^+_{ij} = \epsilon_{ijk} Q^-_k$, $Q^-_i$  (resp.~$Q^+_i$, $P^-_{ij}
=
\epsilon_{ijk} Q^+_k$, $Q^+_i$), where $\epsilon_{ijk}$  is the antisymmetric
tensor and there is a summation over dummy indices.\par
%
%
The components of $\hat H$ can be expressed in terms of such matrices as
\begin{equation}
   H^{(n)} = H'^{(n)} + H''^{(n)}, \qquad H'^{(n)} = Q^+_{n-1,n} Q^-_{n,n-1},
\qquad
   H''^{(n)} = Q^-_{n+1,n} Q^+_{n,n+1},   \label{eq:2.7}
\end{equation}
where $n=0$, 1, 2,~3, and $Q^+_{-1,0} = Q^-_{0,-1} = Q^+_{3,4} = Q^-_{4,3} =
0$.
Apart from some additive constants, $H^{(0)} = Q^+_i Q^-_i$, and
$H^{(3)} = Q^-_i Q^+_i$ are given by~(\ref{eq:2.1}) with $g = \kappa (\kappa -
1)$,
\linebreak $f = \lambda (\lambda - 1)$, and $g = \kappa (\kappa + 1)$,
$f = \lambda (\lambda + 1)$ respectively, while the remaining two components
$H^{(1)}$ and~$H^{(2)}$ have the form of Schr\H odinger operators with matrix
potentials. The operators~(\ref{eq:2.7}) satisfy the intertwining relations
\begin{equation}
  H'^{(n+1)} Q^+_{n,n+1} = Q^+_{n,n+1} H''^{(n)}, \qquad Q^-_{n+1,n} H'^{(n+1)}
=
  H''^{(n)} Q^-_{n+1,n},    \label{eq:2.8}
\end{equation}
and similar relations with $H'^{(n+1)}$ and~$H''^{(n)}$ replaced by $H^{(n+1)}$
and~$H^{(n)}$, respectively. Eq.~(\ref{eq:2.8}) shows that all discrete
eigenvalues
of $H''^{(n)}$ and~$H'^{(n+1)}$ (or, equivalently, $H^{(n)}$ and~$H^{(n+1)}$)
but one
are the same and that their eigenfunctions are connected by the operators
$Q^+_{n,n+1}$, $Q^-_{n+1,n}$.\par
%
%
\section{Generalized Momenta for the CMW Problem without Oscillator Potential}
\setcounter{equation}{0}
In the absence of harmonic oscillator and three-body potentials, i.e., for
$\omega =
f = 0$, the operators~$Q^-_i$, defined in~(\ref{eq:2.3}), can be converted into
covariant derivatives $D_i$ by inserting particle permutation operators
$K_{ij}$,
obeying
\begin{equation}
  K_{ij} = K_{ji} = K_{ij}^{\dagger}, \qquad K_{ij}^2 = 1, \qquad K_{ij} K_{jk}
=
  K_{jk} K_{ki} = K_{ki} K_{ij},   \label{eq:3.1}
\end{equation}
and
\begin{equation}
  K_{ij} x_j = x_i K_{ij}, \qquad K_{ij} x_k = x_k K_{ij},   \label{eq:3.2}
\end{equation}
for all $i\ne j\ne k\ne i$. The operators~\cite{5,6}
\begin{equation}
  D_i = \partial_i - \kappa \sum_{j\ne i} \frac{1}{x_{ij}} K_{ij}, \qquad
  i=1,2,3,    \label{eq:3.3}
\end{equation}
indeed satisfy the relations
\begin{equation}
  K_{ij} D_j = D_i K_{ij}, \qquad K_{ij}D_k = D_k K_{ij} \quad (k\ne i,j),
\qquad
  D_i^{\dagger} = D_i, \qquad [D_i, D_j] = 0,    \label{eq:3.4}
\end{equation}
and
\begin{equation}
  -\sum_i D_i^2 = - \sum_i \partial_i^2 + \sumij \frac{1}{x_{ij}^2} \kappa
(\kappa
  - K_{ij}).      \label{eq:3.5}
\end{equation}
{}From (\ref{eq:3.5}), it results that in those subspaces of Hilbert space
wherein
$K_{ij} = +1$ or $-1$ for any $i$,~$j$, the Calogero Hamiltonian without
oscillator
potential, corresponding to the parameter value $\kappa$ or~$\kappa + 1$
respectively, can be regarded as a free particle Hamiltonian expressed in terms
of
the generalized momenta $\pi_i = - i D_i$.\par
%
%
We now plan to show that this exchange operator formalism can be easily
extended
to the case where the three-body potential is present. For such purpose, we
note
that from~(\ref{eq:3.2})
\begin{equation}
  K_{ij} x_{ij} = - x_{ij} K_{ij}, \qquad K_{ij} x_{jk} = - x_{ki} K_{ij},
\qquad
  K_{ij} R = R K_{ij},    \label{eq:3.6}
\end{equation}
where $i\ne j\ne k\ne i$, and $R = (x_1 + x_2 + x_3)/3$ denotes the
centre-of-mass
coordinate. As the dependence of Hamiltonian~(\ref{eq:2.1}) upon $x_{ij}$
and~$y_{ij}$ is similar, let us introduce some operators~$L_{ij}$ satisfying
properties analogous to~(\ref{eq:3.1}) and~(\ref{eq:3.6}), where in the latter
$y_{ij}$ is substituted for $x_{ij}$,
\begin{eqnarray}
  L_{ij} & = & L_{ji} = L_{ij}^{\dagger}, \qquad L_{ij}^2 = 1, \qquad L_{ij}
L_{jk} =
      L_{jk} L_{ki} = L_{ki} L_{ij}, \nonumber \\
  L_{ij} y_{ij} & = & - y_{ij} L_{ij}, \qquad L_{ij} y_{jk} = - y_{ki} L_{ij},
\qquad
      L_{ij} R = R L_{ij}.   \label{eq:3.7}
\end{eqnarray}
Hence,
\begin{equation}
  L_{ij} x_i = (2R - x_j) L_{ij}, \qquad L_{ij} x_k = (2R - x_k) L_{ij} \quad
  (k\ne i,j),    \label{eq:3.8}
\end{equation}
and $L_{ij}$ can be written as
\begin{equation}
  L_{ij} = K_{ij} I_r = I_r K_{ij}, \qquad I_r = I_r^{\dagger}, \qquad I_r^2 =
1,
  \label{eq:3.9}
\end{equation}
where
\begin{equation}
  I_r x_i = (2R - x_i) I_r, \qquad I_r x_{ij} = - x_{ij} I_r, \qquad I_r y_{ij}
=
   - y_{ij} I_r, \qquad I_r R = R I_r.   \label{eq:3.10}
\end{equation}
\par
%
%
The new operator~$I_r$ is therefore the inversion operator in
relative-coordinate
space. Together with $K_{ij}$, it generates a group of order twelve, which is
the
direct product of the symmetric group $S_3$ and the group of order two $\{1,
I_r\}$,
and is isomorphic to the dihedral group~$D_6$.\par
%
%
For simplicity's sake, from now on we shall work in the centre-of-mass
coordinate
system and therefore set $R\equiv 0$ in (\ref{eq:3.6})--(\ref{eq:3.10}). As a
consequence, any coordinate~$x_i$ may be replaced by $-\sum_{j\ne i} x_j$. This
substitution will play an important role in some of the subsequent
calculations.
\par
%
%
By inserting the operators~$L_{ij}$ into~$Q^-_i$, the operators~(\ref{eq:3.3})
are
generalized into
\begin{equation}
  D_i = \partial_i - \kappa \sum_{j\ne i} \frac{1}{x_{ij}} K_{ij} - \lambda
  \left(\sum_{j\ne i} \frac{1}{y_{ij}} L_{ij} - \sumjk \frac{1}{y_{jk}}
  L_{jk}\right),     \label{eq:3.11}
\end{equation}
whenever $g\ne 0$, and $f\ne 0$. For $g = 0$, and $f\ne 0$, the second term on
the
right-hand side of~(\ref{eq:3.11}) is not present. After some algebra, one
finds that
the operators~(\ref{eq:3.11}) still satisfy  Eq.~(\ref{eq:3.4}), and that in
addition
\begin{equation}
  I_r D_i = - D_i I_r, \qquad L_{ij} D_i = - D_j L_{ij}, \qquad L_{ij} D_k = -
D_k
  L_{ij} \quad (k\ne i,j),   \label{eq:3.12}
\end{equation}
and
\begin{equation}
  -\sum_i D_i^2 = - \sum_i \partial_i^2 + \sumij \frac{1}{x_{ij}^2} \kappa
(\kappa
  - K_{ij}) + 3 \sumij \frac{1}{y_{ij}^2} \lambda (\lambda - L_{ij}).
\label{eq:3.13}
\end{equation}
\par
%
%
The operators ${\cal I}_n = \sum_i \pi_i^{2n}$, where $\pi_i = - i D_i$, $i =
1$,
2,~3, are generalized momenta, commute with one another, and are left invariant
under $D_6$. Hence, their projection in the subspaces of Hilbert space
characterized by $(K_{ij}, L_{ij}) = (1,1)$, $(1,-1)$, $(-1,1)$, or~$(-1,-1)$,
also
commute. In those subspaces,
${\cal I}_1$ becomes the CMW~Hamiltonian corresponding to parameter
values $(\kappa, \lambda)$, $(\kappa, \lambda + 1)$, $(\kappa + 1, \lambda)$,
or
$(\kappa + 1, \lambda + 1)$ respectively, while ${\cal I}_2$ and~${\cal I}_3$
become the integrals of motion of such Hamiltonian.\par
%
%
Note that operators similar to~(\ref{eq:3.11}) were independently derived by
Buchstaber {\em et al}~\cite{14} in a work on generalized Dunkl operators. In
their
approach, use is made of the root system and the Weyl group of the semi-simple
Lie algebra $G_2$, to which the CMW~Hamiltonian is known to be
related~\cite{4}.
Such a Weyl group is just the dihedral group~$D_6$ hereabove considered.\par
%
%
\section{Modified Boson Operators for the CMW~Problem with Oscillator
Potential}
\setcounter{equation}{0}
When the oscillator potential is present in Hamiltonian~(\ref{eq:2.1}), it is
advantageous to introduce modified boson creation and annihilation operators,
defined by
\begin{equation}
  \ac_i = \frac{1}{\sqrt{2\omega}} (\omega x_i - D_i), \qquad
  \ad_i = \left(\ac_i\right)^{\dagger} = \frac{1}{\sqrt{2\omega}} (\omega x_i +
D_i).
  \label{eq:4.1}
\end{equation}
By using~(\ref{eq:3.2}) and~(\ref{eq:3.8}), it can be easily shown that they
satisfy
the commutation relations
\begin{eqnarray}
  \left[\ac_i, \ac_j\right] & = & \left[\ad_i, \ad_j\right] = 0, \nonumber \\
  \left[\ad_i, \ac_i\right] & = & 1 + \kappa \sum_{j\ne i} K_{ij} +
\frac{\lambda}{3}
  \left(\sum_{j\ne i} L_{ij} + 2 \sumjk L_{jk}\right), \nonumber \\
  \left[\ad_i, \ac_j\right] & = & - \kappa K_{ij} + \frac{\lambda}{3} \left(
L_{ij}
      - 2 \sum_{k\ne i,j} (L_{ik} + L_{jk})\right), \qquad i\ne j,
\label{eq:4.2}
\end{eqnarray}
and the exchange relations
\begin{equation}
  K_{ij} a^{(\dagger)}_j = a^{(\dagger)}_i K_{ij}, \qquad K_{ij}
a^{(\dagger)}_k =
  a^{(\dagger)}_k K_{ij} \quad (k\ne i,j), \qquad I_r a^{(\dagger)}_i = -
a^{(\dagger)}_i
  I_r.       \label{eq:4.3}
\end{equation}
\par
%
%
Eqs.~(\ref{eq:3.1}), (\ref{eq:3.9}), (\ref{eq:4.2}), and~(\ref{eq:4.3}) define
a
$D_6$-extended Heisenberg algebra. Whenever $f = 0$ and $g\ne 0$, the terms
proportional to~$\lambda$ on the right-hand side of~(\ref{eq:4.2}) disappear,
so
that the algebra reduces to the $S_3$-extended one considered in
Refs.~\cite{5,6,8}, where $S_3 = \{1, K_{12}, K_{23}, K_{31}, K_{12} K_{23},
K_{23}
K_{12}\}$. Whenever $g = 0$ and $f\ne 0$, the terms proportional to~$\kappa$
are
not present on the right-hand side of~$(\ref{eq:4.2})$, so that the algebra
becomes
a $D_3$-extended Heisenberg algebra, where $D_3 = \{1, L_{12}, L_{23}, L_{31},
L_{12} L_{23}, L_{23} L_{12}\}$. Note that although $D_3$ and $S_3$ are
isomorphic, their action on the creation and annihilation operators is
different,
thus giving rise to different extended algebras.\par
%
%
{}From (\ref{eq:4.1}) and (\ref{eq:3.13}), we obtain
\begin{eqnarray}
  \omega \sum_i \left\{\ac_i, \ad_i\right\} & = & \sum_i \left(- D_i^2 +
\omega^2
       x_i^2\right) \nonumber \\
  & = & - \sum_i \partial_i^2 + \omega^2 \sum_i x_i^2 + \sumij
\frac{1}{x_{ij}^2}
       \kappa (\kappa - K_{ij}) + 3 \sumij \frac{1}{y_{ij}^2} \lambda (\lambda
       - L_{ij}).     \label{eq:4.4}
\end{eqnarray}
Hence, the CMW~Hamiltonian with oscillator potential corresponding to the
parameter values ($\kappa$, $\lambda$), ($\kappa$, $\lambda + 1$),
($\kappa + 1$, $\lambda$), or ($\kappa + 1$, $\lambda + 1$), can be regarded as
a
free modified boson Hamiltonian.\par
%
%
To show the existence of conserved quantities, let us consider the operators
\begin{equation}
  {\cal I}_n = \sum_i h_i^n, \qquad h_i = \case{1}{2} \left\{\ac_i,
\ad_i\right\},
  \qquad n=1,2,3.    \label{eq:4.5}
\end{equation}
{}From the defining relations of the $D_6$-extended Heisenberg algebra, it can
be
shown that the operators~$h_i$ obey the commutation relations
\begin{equation}
  [h_i, h_j] = \case{1}{4} \left(\kappa^2 + \lambda^2 - 2 \kappa \lambda
I_r\right)
  \sum_{k\ne i,j} \left(K_{ij} - K_{ik}\right) K_{jk}, \qquad i\ne j.
\label{eq:4.6}
\end{equation}
It results that ${\cal I}_1$ commutes with any $h_i$, hence with~${\cal I}_2$
and~${\cal I}_3$.\par
%
%
{}From (\ref{eq:4.6}), it also follows that
\begin{eqnarray}
  [{\cal I}_2, h_i] & = & \sum_j \left(h_j [h_j, h_i] + [h_j, h_i] h_j\right)
\nonumber
       \\
  & = & \case{1}{4} \left(\kappa^2 + \lambda^2 - 2 \kappa \lambda I_r\right)
       \sum_{\scriptstyle j,k \atop \scriptstyle i\ne j< k\ne i} \left[(h_i
-h_j) K_{ij}
       + (h_i - h_k) K_{ik}\right] K_{jk},   \label{eq:4.7}
\end{eqnarray}
from which we obtain
\begin{eqnarray}
  \left[{\cal I}_2, h_i^3\right] & = & [{\cal I}_2, h_i] h_i^2 + h_i [{\cal
I}_2, h_i] h_i
        + h_i^2 [{\cal I}_2, h_i] \nonumber \\
  & = & \case{1}{4} \left(\kappa^2 + \lambda^2 - 2 \kappa \lambda I_r\right)
       \sum_{\scriptstyle j,k \atop \scriptstyle i\ne j< k\ne i}
\left[\left(h_i^3
       -h_j^3\right) K_{ij} + \left(h_i^3 - h_k^3\right) K_{ik}\right] K_{jk}.
       \label{eq:4.8}
\end{eqnarray}
It is then straightforward to prove that ${\cal I}_2$ also commutes with
${\cal I}_3$.\par
%
%
So we did obtain three mutually commuting conserved quantities~${\cal I}_n$,
$n=1$, 2,~3. Since the latter are invariant under $D_6$, their projections in
the
subspaces characterized by~$K_{ij}$ and~$L_{ij}$ equal to~$+1$ or~$-1$ do still
commute with one another. In those subspaces, ${\cal I}_1$ becomes proportional
to the corresponding CMW~Hamiltonian, while ${\cal I}_2$ and~${\cal I}_3$
become the integrals of motion of the latter.\par
%
%
\section{Conclusion}
In the present letter, we did show that the exchange operator formalism
previously
introduced for the Calogero problem can be extended to the three-body CMW~one,
thus providing us with an easy proof of complete quantum integrability for the
latter. In the analysis of the problem, we were led to consider a
$D_6$-extended
Heisenberg algebra whenever two and three-body interactions are both present,
and
a $D_3$-extended one whenever only the latter is taken into account.\par
%
%
Whether a similar approach can be used for some other Hamiltonians containing
three-body interactions remains an open question, to which we hope to come back
in a near future.\par
%
%
\section*{Acknowledgments}
The author would like to thank T.~Brzezi\'nski for some helpful
discussions.\par
\newpage
%
%


\begin{thebibliography}{99}

\bibitem{1} F. Calogero, {\em J. Math. Phys.} \textbf{10}, 2191 (1969).

\bibitem{2} J. Wolfes, {\em J. Math. Phys.} \textbf{15}, 1420 (1974).

\bibitem{3} F. Calogero and C. Marchioro, {\em J. Math. Phys.} \textbf{15},
1425
(1974).

\bibitem{4} M. A. Olshanetsky and A. M. Perelomov, {\em Phys. Rep.}
\textbf{71},
313 (1981), \textbf{94}, 313 (1983).

\bibitem{5} L. Brink, T. H. Hansson and M. A. Vasiliev, {\em Phys. Lett.}
\textbf{B286}, 109 (1992).

\bibitem{6} A. P. Polychronakos, {\em Phys. Rev. Lett.} \textbf{69}, 703
(1992).

\bibitem{7} C. F. Dunkl, {\em Trans. Am. Math. Soc.} \textbf{311}, 167 (1989).

\bibitem{8} L. Brink and M. A. Vasiliev, {\em Mod. Phys. Lett.} \textbf{A8},
3585
(1993).

\bibitem{9} T. Brzezi\'nski, I. L. Egusquiza and A. J. Macfarlane, {\em Phys.
Lett.}
\textbf{B311}, 202 (1993).

\bibitem{10} J. M. Leinaas and J. Myrheim, {\em Phys. Rev.} \textbf{B37}, 9286
(1988); A. P. Polychronakos, {\em Nucl. Phys.} \textbf{B324}, 597 (1989).

\bibitem{11} D. Bernard, M. Gaudin, F. D. M. Haldane and V. Pasquier, {\em J.
Phys.}
\textbf{A26}, 5219 (1993).

\bibitem{12} A. Khare and R. K. Bhaduri, {\em J. Phys.} \textbf{A27}, 2213
(1994).

\bibitem{13} A. A. Andrianov, N. V. Borisov and M. V. Ioffe, {\em Phys. Lett.}
\textbf{A105}, 19 (1984); A. A. Andrianov, N. V. Borisov, M. I. Eides and M. V.
Ioffe,
{\em Phys. Lett.} \textbf{A109}, 143 (1985).

\bibitem{14} V. M. Buchstaber, G. Felder and A. P. Veselov, ``Elliptic Dunkl
operators,
root systems, and functional equations'', preprint hep-th/9403178, March 1994.

\end {thebibliography}
\end{document}